# High-Efficiency All-Dielectric Metasurfaces for Ultra-Compact Beam Manipulation in Transmission Mode

*Mikhail I. Shalaev[1], Jingbo Sun[1], Alexander Tsukernik[2], Apra Pandey[3], Kirill Nikolskiy[4], Natalia M. Litchinitser[1]\**

1 *Department of Electrical Engineering, University at Buffalo, The State University of New York, Buffalo, NY 14260, USA*

2 *Toronto Nanofabrication Centre, University of Toronto, Toronto, ON M5S 3G4, Canada*

3 *CST of America, Inc., San Mateo, CA 94404, USA*

4 *Physics Department, M.V. Lomonosov Moscow State University, Moscow 119991, Russia*

ABSTRACT: Metasurfaces are two-dimensional optical structures enabling complete control of the amplitude, phase, and polarization of light. Unlike plasmonic metasurfaces, planar silicon structures facilitate high transmission, low losses and compatibility with existing semiconductor technologies. Here, we report an experimental demonstration of high-efficiency polarization-sensitive dielectric metasurfaces with full $2\pi$ phase control in transmission mode at telecommunication wavelengths. Such silicon metasurfaces are poised to enable a versatile optical platform the realization of all-optical circuitry on a chip.

TEXT. Metasurfaces are two-dimensional artificial materials with thicknesses much smaller than incident light wavelength, allowing complete control of the phase, amplitude, and polarization of light beams.[1-25] Compared to conventional optical elements, which rely on long propagation distances, these devices facilitate strong light-matter interaction on a subwavelength scale, allowing abrupt changes of beam parameters. Metasurfaces, unlike their three-dimensional analogues, metamaterials, do not require complicated fabrication techniques and can be produced in one lithographical step, which makes them very promising for integration on a photonic chip and well suited for mass production. To date, a majority of the studies has been focused on metal-dielectric structures, which have relatively low efficiency due to orthogonal polarizations coupling and nonradiative Ohmic losses in metals.[1, 2, 16] Many potential applications of metasurfaces, such as beam steering, lensing, holography or structured light generation, would require full $2\pi$ phase control. However, it was shown that in the case of isolated single electric or magnetic resonance, only $\pi$ phase shift is possible.[1, 2, 26, 27] While such phase manipulation can be obtained through cascading of multiple functional layers or by operating in reflection mode[1, 2, 25, 28], metasurfaces realized using these approaches are not easily integratable on a chip and moreover, may not be compatible with contemporary semiconductor industry technologies or may not be well-suited for mass production.

Recently, it has been shown that high-refractive index nanoparticles embedded in a low-index surrounding medium can be designed such that both magnetic and electric resonant responses occur in the same frequency range.[29-33] It is the presence of both electric and magnetic resonances at the same frequencies allows a $2\pi$ phase control in a single-layer all-dielectric structure. It was shown that silicon nanostructures, having a relatively high refractive index at telecommunication wavelength, can be optimized to possess overlapping electric and magnetic dipole resonances in the same frequency range.[26, 27, 34] Also, silicon is the most commonly used material in the semiconductor industry, which makes it an ideal platform for high-efficiency metasurfaces for near-infrared (NIR) wavelength and future integration on optical chip.

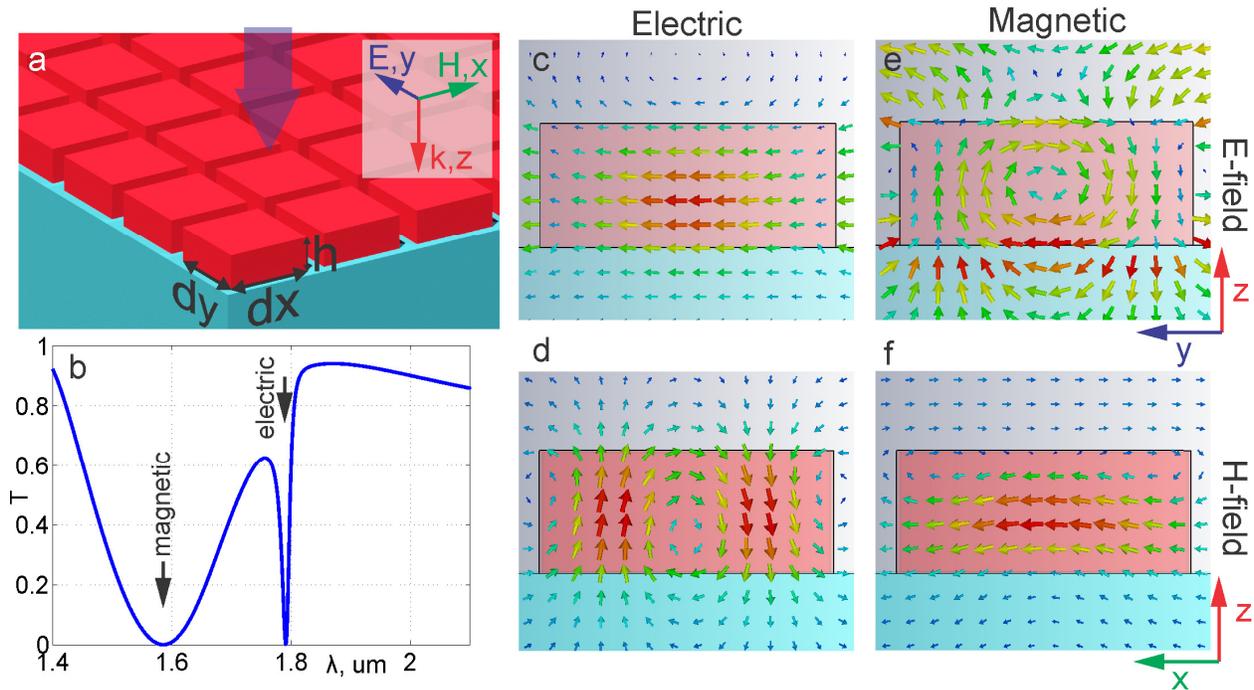

**Figure 1.** (a) Schematics of an infinite array of silicon nanoblocks metasurface on top of bulk-fused silica substrate. Nanoblocks height *h*=270nm, dimensions along *x* and *y* axes *dx*=*dy*=650nm, lattice constant *a*=800nm. (b) Transmission spectrum of metasurfaces shows two dips for magnetic ($\lambda_e$=1.6$\mu m$) and electric ($\lambda_m$=1.8$\mu m$) resonances. (c, d) Electric field enhancement at the center of silicon nanoblocks and vortex-like magnetic field distribution indicates electric resonance, while the opposite holds for magnetic resonance (e, f) with *H*-field maximum at the center and twisted *E*-field around it.

First, we consider light propagation through an infinite array of high refractive index polysilicon nanoblocks (refractive index *n*=3.67 with height *h*=270*nm* and dimensions *dx*=*dy*=650*nm*, on top of semi-infinite fused silica substrate (refractive index $n_s$=1.45), lattice constant *a*=800*nm* (Figure 1*a*). We used commercially available CST MICROWAVE STUDIO software to perform numerical simulation in frequency domain; the incident electromagnetic field was assumed to be a plane wave, propagating along *z* axis, with electric and magnetic fields polarized along *y* and *x* axes, correspondingly. The transmission spectrum, shown in Figure 1b, clearly shows two dips corresponding to resonant interaction with metasurfaces and, as a result, near unity reflection, while the structure was assumed to be lossless. Electric and magnetic field distributions in the unit cell cross-section show magnetic resonance behavior for wavelength around $\lambda_e$=1.6*μm*, with *E*-field concentrated at the center of nanoblocks and vortex-like *H*-field distribution around electric field (Figure 1c, d). The opposite holds for electric resonance around $\lambda_m$=1.8*μm*, with magnetic field maximum at the center in the *y-z* plane and vortex-like electric field around in the *x-z* plane (Figure 1e, f). In this example electric and magnetic resonances are well-separated and can be easily distinguished. However, as the block size decreases, the spacing between the resonances decreases such that two resonances shift toward the wavelength of interest. Once the resonances overlap, enabling the impedance matching, nearly 100% transmission with full 2π phase control can be achieved. We calculate transmitted beam phase as a function of silicon block size along *x* and *y* axes. Note that the possibility of varying the dimensions of the nanoblocks in both *x* and *y* directions adds additional degrees of freedom and allows the design of polarization-dependent metasurfaces; polarization-independent design is also possible and was theoretically studied in Reference 27. In this work, the wavelength of interest $\lambda_0$=1.55*μm*, lattice constant *a*=800*nm*, and silicon nanoblock height *h*=270*nm*. Figure 2 shows the results of numerical simulation for phase and corresponding transmittance as a function of *dx* and *dy*. As one of the advantages of the all-dielectric metasurface design is the possibility of high transmission through the structure, we use the results of these numerical simulations to simultaneously enable 0 to $2\pi$ variation of the phase as well as transmission around 80%. Also, we limit the dimensions of the nanoblocks to be less than 750*nm*. This requirement is due to fabrication constraints as we fixed the lattice constant to be 800*nm*. The results shown in Figure 2 are general rather than application specific, and therefore, they can be used to design a number of functionalities.

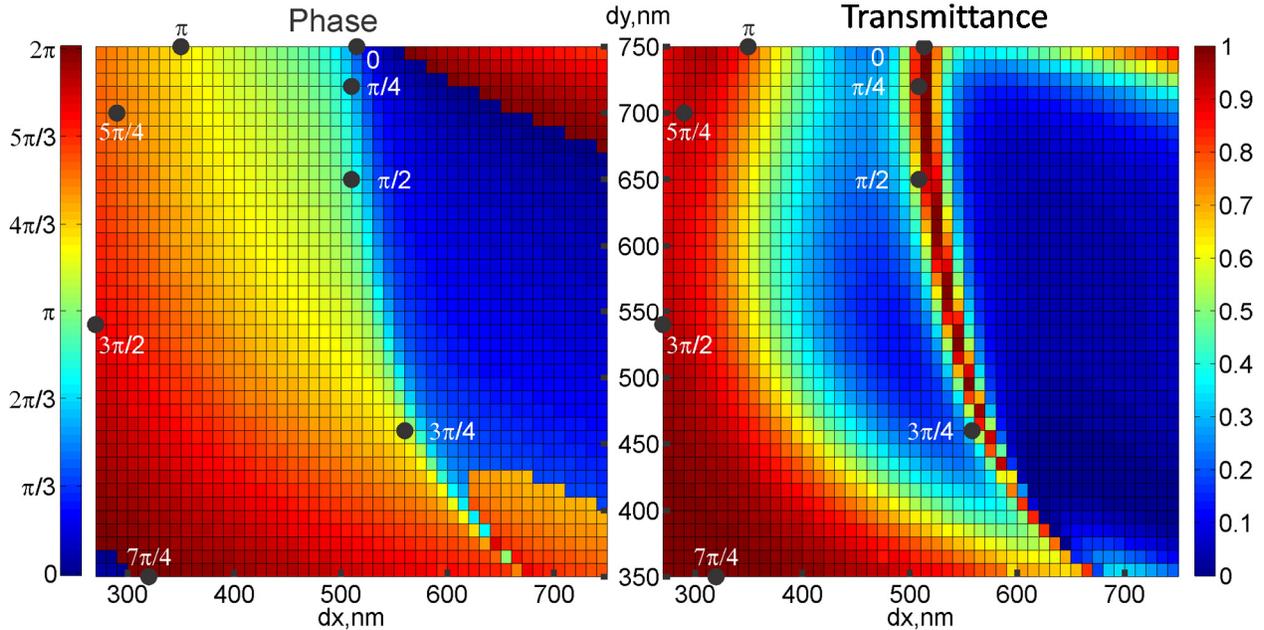

**Figure 2.** (a) Transmitted light phase variation for nanoblocks on silicon dioxide substrate with lattice constant and (b) normalized transmittance as a function of silicon block size *dx* and *dy*.

We choose eight discrete nanoblocks with $\pi/4$ increments to cover 0-to-$2\pi$ phase to provide full phase control of the wavefront. Table 1 shows transmitted light phase for nanoblocks dimensions and corresponding phase shift.

**Table 1.** Optimized nanoblocks dimensions.

| Phase | 0 | $\pi/4$ | $\pi/2$ | $3\pi/4$ | $\pi$ | $5\pi/4$ | $3\pi/2$ | $7\pi/4$ |
|---|---|---|---|---|---|---|---|---|
| *dx*, nm | 515 | 510 | 510 | 560 | 350 | 290 | 270 | 320 |
| *dy*, nm | 750 | 720 | 650 | 460 | 750 | 700 | 540 | 350 |

For experimental verification of our numerical simulations, we deposited 270*nm* of silicon with low pressure physical vapor deposition (LPCVD) on top of fused quartz wafers. Fabrication process is shown in Figure 3. The refractive index of deposited silicon was measured to be *n*=3.67 at 1.55*μm* using spectroscopic ellipsometry. We used standard electron-beam lithography (EBL) with ZEP520A resist; due to the strong charging effect of nonconductive $SiO_2$ substrate, 20nm thick charge dissipation Cr layer was deposited on the e-beam resist. The EBL pattern writing was followed by Cr etching with commercially available Cr etchant and development in ZED-N50. Then deep reactive ion etching (DRIE) was performed in $C_4F_8$ and $SF_6$ gases, followed by resist removing in Remover 1165 at 80°C for 1hour and 5minutes $O_2$ plasma for removal of resist residue.

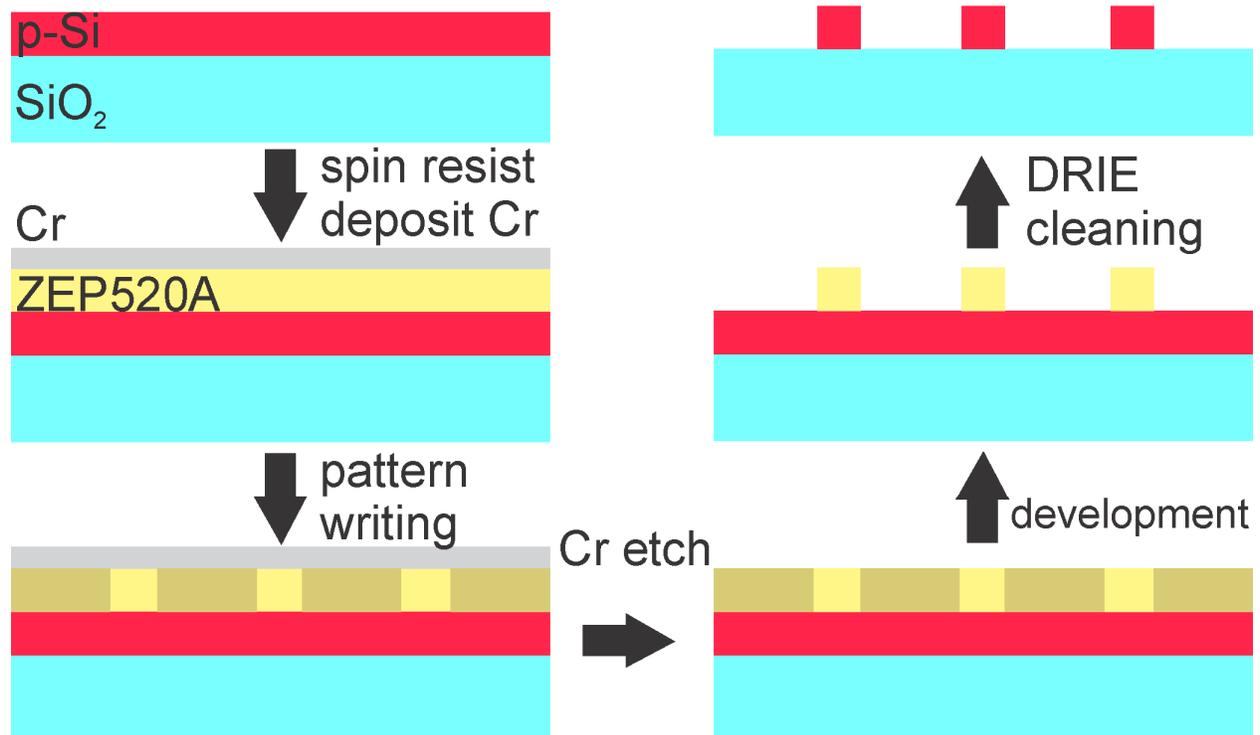

**Figure 3.** Fabrication process for silicon-based dielectric metasurfaces. Polycrystalline silicon was deposited on top of fused silica wafers by low-pressure chemical vapor deposition (LPCVD), followed by spin coating of e-beam resist ZEP520A and deposition of 20nm charge dissipation Cr layer. Than pattern was written by electron beam lithography (EBL) and Cr layer was etched by ceric ammonium nitrate-based etchant. Samples were developed in ZED-N50 followed by deep reactive ion etching (DRIE), removal of resist and $O_2$ plasma cleaning.

We experimentally demonstrated high-efficiency full-phase manipulation with dielectric metasurfaces at near-infrared wavelength enabling two practical functionalities: optical beam steering and conversion of a conventional Gaussian beam into a beam with an orbital angular momentum (OAM) or a vortex beam. In both cases, dramatic changes on light propagation take place within a distance of less than $\lambda/5$, opening new possibilities of light control with semiconductor-industry-compatible materials and an easy fabrication procedure.

Figure 4a, b shows schematics and scanning electron microscopy (SEM) images for a fabricated beam deflecting metasurface with 96×96μm total size, unit cell is shown in inset. The metasurface contains eight nanoblocks from table 1, where each is responsible for a phase shift from 0 to $2\pi$ with $\pi/4$ increments. We consider light propagation from isotropic medium 1 to the medium 3 through bulk substrate with the metasurface on top as shown in the inset in Figure 4c. Deflection angle for a beam can be calculated with the following equation[35] $\theta_3 = \sin^{-1}[(n_1 \sin\theta_1 + \lambda_0/\Gamma)/n_3]$, where $n_1$ and $n_3$ are refractive indices for media 1 and 3, $\theta_1$ - incidence angle, $\lambda_0$ - free-space wavelength of light, $\Gamma$ - periodicity of the structure. We consider normal incidence of a light beam on the metasurface with the period of the structure $\Gamma=6.4\mu m$ fabricated on the substrate surrounded by air. In this case the angle of refraction is $\theta_3 \approx 14°$. We consider plane wave light propagation through infinite two-dimensional periodic array of silicon

nanoblocks. Numerical simulations confirm the theoretically calculated refraction angle, as shown in Figure 3c.

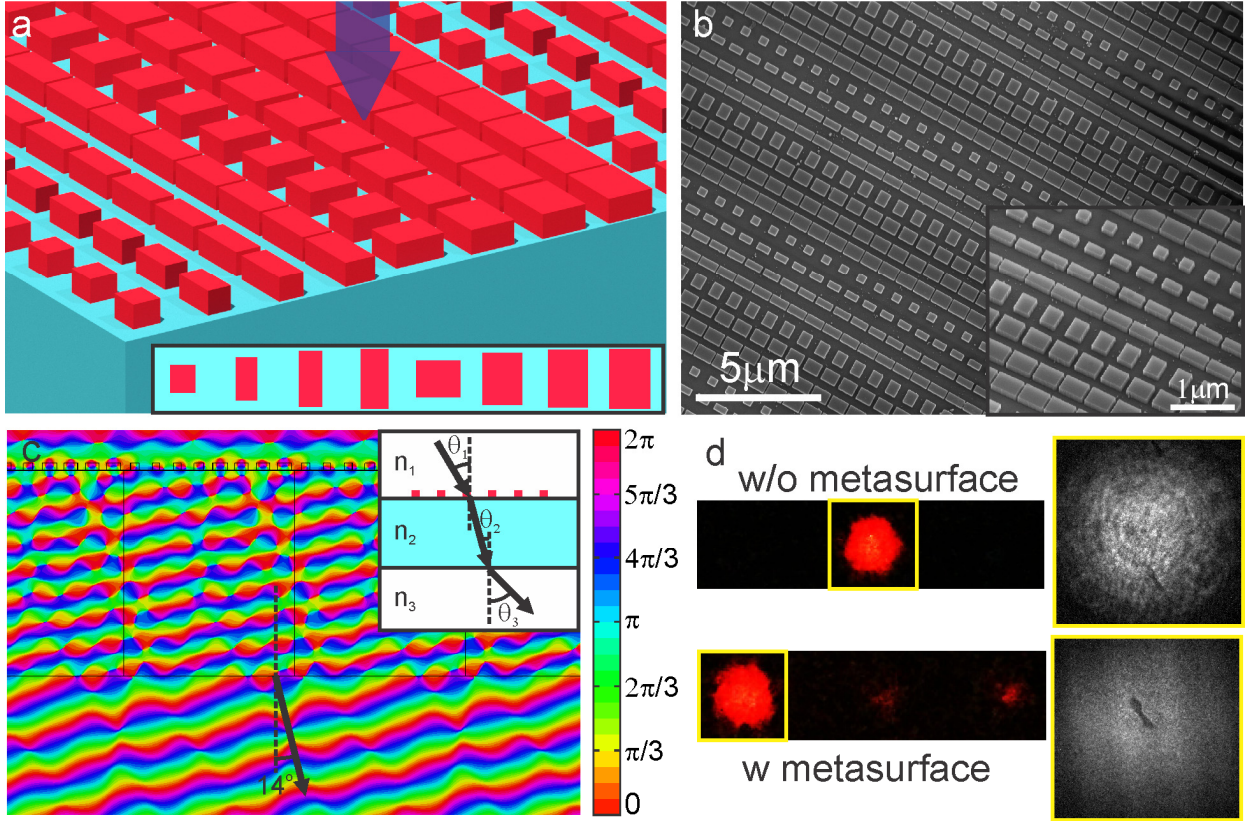

**Figure 4.** (a) Schematics of the proposed beam deflector, with the unit cell (shown in the inset) containing eight nanoblocks, each responsible for the phase shift from 0 to $2\pi$ with $\pi/4$ increments. (b) Scanning electron microscopy image of fabricated metasurfaces with 96×96$\mu m$ total size; zoomed-in picture is shown in the inset. (c) Phase of plane wave propagating through the metasurfaces on $SiO_2$ substrate showing refracted beam angle around 14 degrees. (d) Beam position without (left top) and with (left bottom) metasurface; most of the beam power is refracted to the left side with deflection angle 13.1° and transmission of ~36%. Input and output beam photographs are shown in the inset (right).

We used a diode laser as a light source to perform measurements for beam steering using a NIR detector card to determine deflected beam position; schematic of experimental setup is shown in Figure 5. The card converts invisible IR signal to visible region; thus, deflected beam position was captured by visible light camera. The laser beam was focused by lens to the spot with approximately 45$\mu m$ waist. Inset on the left in Figure 4d shows the results of beam-position measurement without (top) and with (bottom) metasurface. As can be seen, most of the beam power is refracted to the left side of the screen while other diffraction orders can be also seen. The presence of other diffraction orders with much smaller intensities compared to the main beam can be caused by the fact that the unit cell size is comparable to wavelength of light[36] by imperfections in fabrication process and violation of local periodicity assumed in our design. The refraction angle for the main beam is measured to be 13.1°, which is close to theoretical and numerical simulations' predictions. Insets on the right show NIR camera images of input and output beams with measured transmission power normalized on input power to be around 36%.

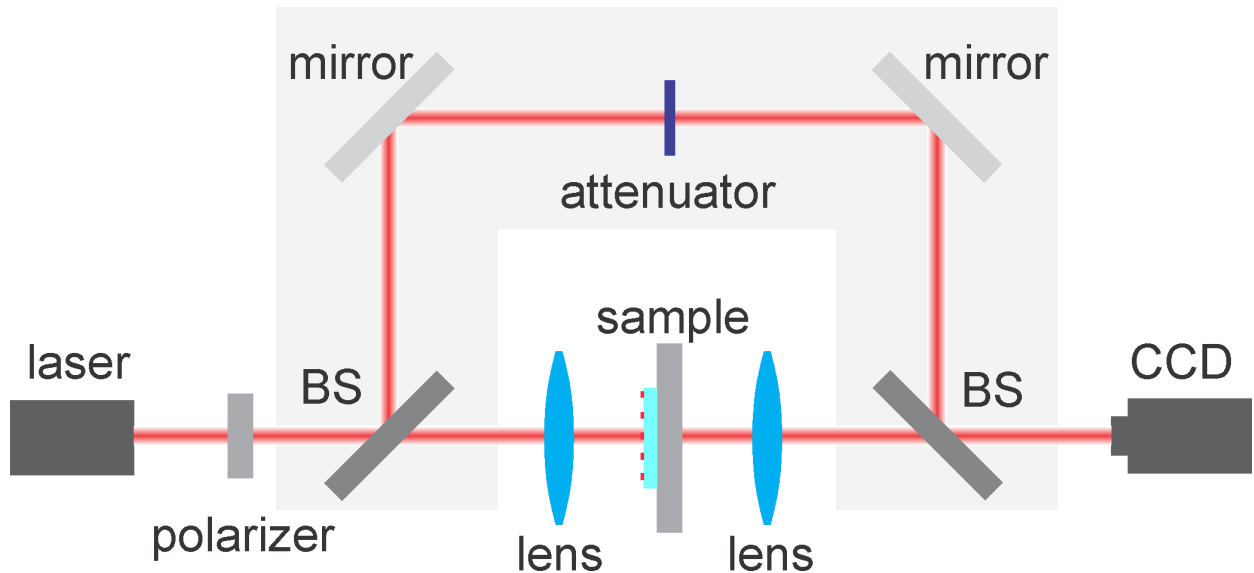

**Figure 5.** Schematics of experimental setup for measurement of fabricated metasurfaces. Diode NIR laser was used as a light source followed by light polarizer, beam was focused to the sample

by lens, then collimated and image was captured with CCD camera. Setup for vortex beam interference is shown in gray background; it contains attenuator, two beam splitters, and mirrors.

Finally, to demonstrate flexibility of designed dielectric nanoblocks for phase control, we fabricated a spatial light modulator that converts a regular Gaussian laser beam into an OAM beam, or vortex. Such structured light beams have a potential for applications ranging from quantum information processing and high-dimensional communication systems to optical manipulation on nanoscale[11, 23, 24, 37, 38]. Schematics of fabricated metasurface for twisted beam generation is shown in Figure 6a, which contains eight sectors for the particular 0-to-$2\pi$ phase shift. Optical vortices, unlike conventional Gaussian beams, possess a donut-shape intensity profile and helical wavefront with phase change from 0 to $2\pi$ in cross-section. Fabricated beam convertor SEM image is shown in Figure 6b, where the total size of the sample is 96×96μm. We performed numerical simulations in CST MICROWAVE STUDIO software package for reduced size model with 67.2×67.2μm size to reduce calculation time and memory consumption. Figures 6c and 6d show the results of numerical simulations corresponding to the normalized transmitted beam intensity and phase at distance of $0.5\lambda_0$ from metasurface, respectively. Corresponding experimentally measured intensity profile is shown in Figure 6e. In order to prove the presence of the helical wavefront, we also performed the interference experiments for both cases of parallel and tilted Gaussian beam interfering with the beam obtained from the metasurface. Resulting spiral- and fork-like intensity profiles are shown in Figure 6f and 6g, respectively. Transmitted power normalized to input power was measured to be 45%.

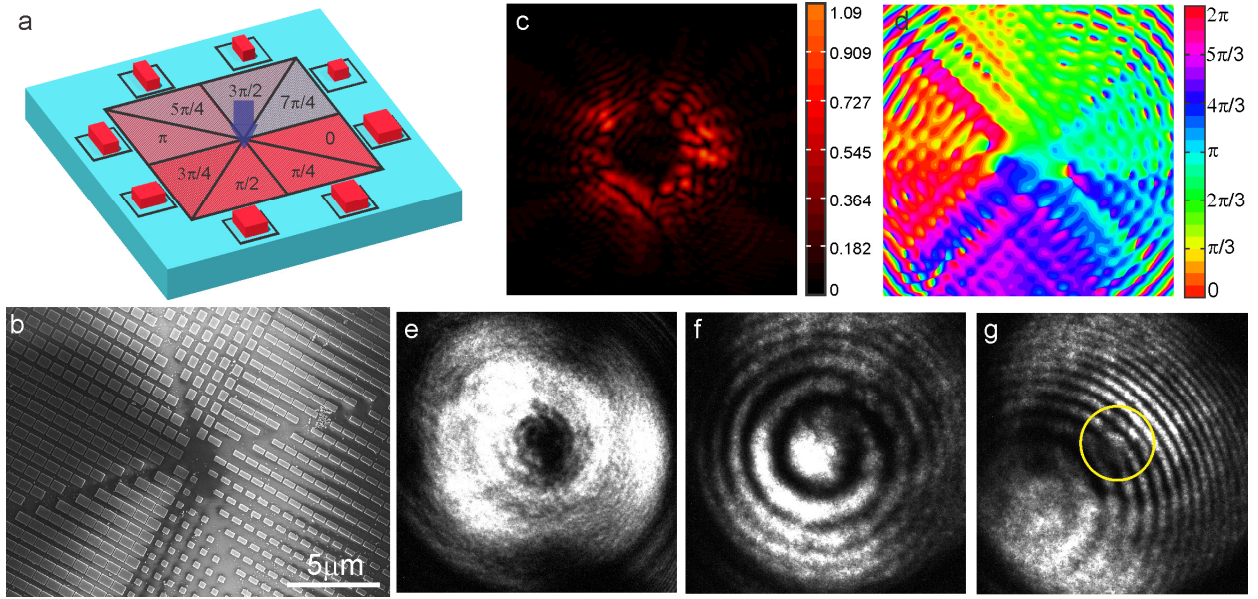

**Figure 6.** (a) Schematics and (b) scanning electron microscopy image for optical vortex beam convertor with eight sectors with nanoblocks from table 1, unit cells for each sector of metasurface are shown. Each sector introduce an additional $\pi/4$ phase shift, thus covering 0 to $2\pi$ phase change. (c, d) Numerically calculated normalized amplitude and phase distribution. (e) Intensity distribution for measured output vortex beam in form of donut shape. (f, g) Vortex and Gaussian beams interference experiment results showing spiral-shaped and fork-like intensity distribution.

In summary, we have experimentally demonstrated an all-dielectric resonant metasurface with full 0-to-$2\pi$ phase control at NIR wavelength. We designed and fabricated high-efficiency beam deflector and light convertor for generating optical vortex beam, carrying an OAM. In addition to 0 to pi phase control enabled by a single magnetic or electric resonance, overlapping of magnetic and electric resonances in the frequency domain enable an additional phase control with optical impedance matching and, as a result, high efficiency in transmission mode with full $2\pi$ phase manipulation. Fabricated devices made with silicon and, in sharp contrast to plasmonic metasurfaces, are compatible with complementary metal–oxide–semiconductor technology. Demonstrated metasurfaces have relatively high transmission coefficients of 36% for beam deflector and 45% for vortex beam generator. These devices can be fabricated in one lithographical step, alleviating the need for cascading or working in reflection mode to achieve full phase control. Silicon-based metasurfaces can be used for fabrication of miniaturized high-efficiency optical components for NIR photonics, such as flat lenses, beam deflectors, anti-reflection coatings and phase modulators. Elimination of metals and, as a result, no Ohmic losses, compared to plasmonic counterparts, and not requiring cross-polarized field interaction to cover full phase, makes them well-suited for integration on optical chip and for large-scale production.

The authors would like to thank Prof. Gennady Shvets and his research group at the University of Texas at Austin for very useful comments on the manuscript. This work was supported by the U.S. Army Research Office Award # W911NF-11-1-0333.

## AUTHOR INFORMATION

**Corresponding Author**

*To whom correspondence should be addressed: natashal@buffalo.edu

**Author Contributions**

The manuscript was prepared in a collaborative effort of all authors. All authors have given approval to the final version of the manuscript. N.M.L. and M.I.S. proposed the idea developed in this work. M.I.S., J.S., A.P. and K.N. made the design and performed the numerical simulations. M.I.S. and A.T. did the fabrication of the samples. J.S. and M.I.S. performed the optical characterization of the sample. N.M.L. and M.I.S. wrote the paper. N.M.L. supervised this work.

**Funding Sources**

U.S. Army Research Office Award # W911NF-11-1-0333

**Notes**

The authors declare no competing financial interest.

ACKNOWLEDGMENT

This work was supported by the U.S. Army Research Office Award # W911NF-11-1-0333.